\documentclass[aps,twocolumn]{revtex4}
\usepackage{epsfig}

\newcommand{\beq}{\begin{equation}}
\newcommand{\eeq}{\end{equation}}
\newcommand{\bea}{\begin{eqnarray}}
\newcommand{\eea}{\end{eqnarray}}
\begin{document}
\title{ Competition between Diffusion and Fragmentation:\\An Important
Evolutionary Process of Nature}
\author{Jesper Ferkinghoff-Borg, Mogens H. Jensen, Joachim Mathiesen,
Poul Olesen and Kim Sneppen\footnote{email: borg,mhjensen,mathies,polesen,sneppen@nbi.dk} }
\affiliation{Niels Bohr Institute,
Blegdamsvej 17, DK-2100 Copenhagen {\O}, Denmark }
\date{\today}
\begin{abstract}
We investigate systems of nature where the common physical processes
diffusion and fragmentation compete. 
We derive a rate equation for the size distribution
of fragments. The equation leads to a third order differential
equation which we solve exactly in terms of Bessel functions. The
stationary state is a universal Bessel distribution described by one
parameter, which fits perfectly experimental data from two very
different system of nature, namely the distribution of ice crystal sizes
from the Greenland ice
sheet and the length distribution of $\alpha$-helices in proteins.
\end{abstract}

\maketitle


A diffusion process is one of the most important and common
physical phenomena of nature. Coherent structures, like crystals
and structural elements of bio-molecules may shrink or grow
gradually and randomly in a way that resembles a diffusive motion
of their boundaries.
On the one hand, inhomogeneities in a particular structure may
disappear or be resolved into neighboring structures due to
diffusion.
On the other hand, diffusion typically tends to make the
largest of the coherent structures larger. In contrast to this gradual
modification of individual fragments,
one might in addition
encounter a completely different
and abrupt physical phenomena, namely that of fragmentation.
Fragmentation occurs in for example growing ice crystals, when
the diffusively growing
crystal is subjected to stresses that can cause a piece to
break off, thus leading to two individual crystals, each of them
continuing the competitive process between
diffusion and fragmentation.

We believe that such interplay between gradual diffusion and
rare but drastic fragmentation is a
very common phenomenon in nature.
In this letter we derive a general
rate equation for the dynamical evolution of the density of
fragments of a given size. The rate equation consists of a local
diffusion term and non-local fragmentation terms. Differentiating
once we obtain a third order differential equation in the size
distribution of fragments. We solve the equation exactly by means
of an eigenvalue expansion in terms of Bessel functions. The
different eigenvalues correspond to different decay times such
that in the infinite time limit we obtain the stationary
distribution, corresponding to zero eigenvalue. This distribution
is characterized by one parameter (the ratio between the diffusion
and the fragmentation constants) and constitutes a universal
stationary distribution for these competing processes. We apply
experimental data from two very different physical processes,
the size distribution of ice crystals in the Greenland ice sheet
\cite{is} and the length distribution of $\alpha$-helices in proteins of
low homology. In both cases we find that the universal distribution
fits the experimental data perfectly. In addition, for the ice
crystals, we are able to map out the entire dynamics of ice
crystals 2000 years back in time \cite{is}.

In order to clarify the description, we consider a one-dimensional
simplification of the process in which the sizes of for instance ice-crystal
grains are projected onto the coordinate $\tilde x$.
The mechanism we propose then is described in terms of the density
$N(\tilde x,\tilde t)$
of ``objects'' (e.g. ice crystals or $\alpha$-helices in proteins)
of a given length $\tilde x$ at a given time $\tilde t$. At a given time it is
possible to increase or decrease $N(\tilde x,\tilde t)$ by diffusion characterized by
the diffusion constant $D$. Also, $N(\tilde x,\tilde t)$ can be altered by fragmentation of
the ``objects'' characterized by a fragmentation constant $f$, defined as the
average number of break-ups in a time interval $d\tilde t$ over a length $L$. Thus,
the number of fragments is given by $fLd\tilde t$. The constants $D$ and $f$ can
depend on temperature and on quantities which are characteristics of the
``objects''. The resulting equation is
\begin{equation}
\frac{\partial N(\tilde x,\tilde t)}{\partial \tilde
  t}=D\frac{\partial^2 N(\tilde x,\tilde t)}{\partial \tilde x^2}
-f\tilde xN(\tilde x,\tilde t)+2f\int_{\tilde x}^\infty d\tilde x'
  N(\tilde x',\tilde t).
\label{first}
\end{equation}
Note that by this equation, the total ``mass" of the fragments
$\int_0^\infty \tilde x N(\tilde x,\tilde t) d\tilde x$ is conserved at any time in the evolution.
The first term on the right hand side represents diffusion, the second term
gives the fragmentation of objects with length $\tilde x$, and the last term
is the contribution from fragmentation of objects larger than $\tilde x$. The
factor of two in the last term follows formally from the condition of
mass conservation, and results from the fact that there are two
distinct ways to cut an object of length $\tilde x+\tilde x'$ into segments of
length $\tilde x$ and $\tilde x'$. 
Similar approach has been applied
in \cite{flyvbjerg} for the stochastic dynamics of microtubules \cite{flyvnote}.

We introduce the dimensionless variables
$x=\tilde x/\tilde x_0$ and $t=\tilde t/\tilde t_0$, where
\begin{equation}
\tilde t_0=(Df^2)^{-1/3}~~{\rm and}~~\tilde x_0=(D/f)^{1/3}.
\label{eq2}
\end{equation}
This amounts to considering eq. (\ref{first}) with $D=f=1$, which we shall do
in the following. We now differentiate eq. (\ref{first}) with respect to $x$
and obtain
\begin{equation}
\partial_t\partial_x N(x,t)=\partial_x^3 N(x,t)-3N(x,t)-x\partial_xN(x,t).
\label{1}
\end{equation}
This equation can be solved by separation, using the boundary condition
that $N(x,t)\rightarrow 0$ for $x\rightarrow \infty$. The result can be
expressed in terms of the Airy function
\begin{equation}
A(x)=\int_0^\infty~dk~\cos \left(kx+\frac{k^3}{3}\right),
\label{A}
\end{equation}
which in turn can be expressed in terms of Bessel functions \cite{watson},
\begin{eqnarray}
&&A(x)=\sqrt{\frac{x}{3}}~K_{\frac{1}{3}}\left(\frac{2x^{3/2}}{3}\right)~{\rm
for}~x>0\nonumber \\
&&=\frac{\pi}{3}\sqrt{|x|}\left[J_{\frac{1}{3}}\left(\frac{2|x|^{3/2}}{3}
\right)+
J_{-\frac{1}{3}}\left(\frac{2|x|^{3/2}}{3}\right)\right]~\nonumber \\
&&{\rm for}~x<0.
\label{bessel}
\end{eqnarray}
The solution of (\ref{1}) is then
\begin{equation}
N(x,t)=\sum_n ~C_ne^{\lambda_n t}B(x+\lambda_n),~B(x)=-\partial^2_x
A(x),
\label{solution}
\end{equation}
where the sum may be an integral in case the eigenvalues $\lambda_n$ are
in a continuous range. The function $B(x)$ is thus related to the Airy
function by the second eq. (\ref{solution}). One further has
\begin{equation}
B(x)=-xA(x).
\label{B-A}
\end{equation}
This follows from the second eq. (\ref{solution}) by performing two
differentiations of the expressions (\ref{bessel}) and by use of standard
Bessel function recursion relations. For a discussion of the Airy function we
refer to \cite{watson}.
\begin{figure}[htbp]
  \begin{center}
    \epsfig{width=.4\textwidth,file=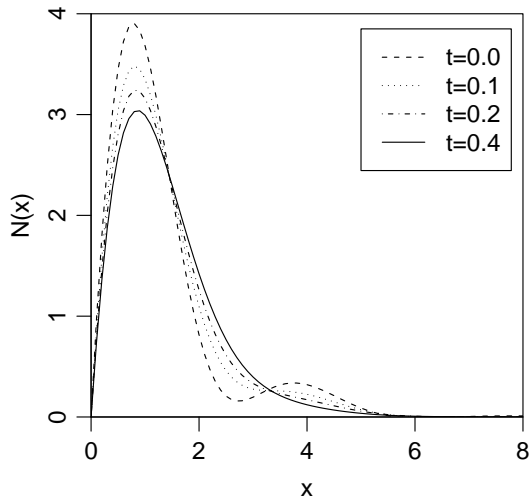}
  \end{center}
  \caption{An example showing
the time development of $N(x,t)$ with a secondary peak, which ultimately
disappears. The specific values of the expansion parameters $C_n$ are
$(C_0,C_1,C_2 \ldots)=(-7,-.2,-.2,.05,-.035)$ .}
  \label{fig:3}
\end{figure}

We now impose the boundary condition that there should be no 
objects of size $x=0$,
\begin{equation}
N(0,t)=0.
\label{boundary}
\end{equation}
To implement the requirement (\ref{boundary}) on the solution (\ref{solution})
we thus need
that the eigenvalues $\lambda_n$ are the zeros of $B(x)$. From eq.
(\ref{B-A}) we see that $x=0$ is a zero. For $x>0$ there are no zeros, whereas
for $x<0$ there exist an infinite number of zeros given as the solution of
the equation
\begin{equation}
J_{\frac{1}{3}}\left(\frac{2|\lambda_n|^{3/2}}{3}
\right)+
J_{-\frac{1}{3}}\left(\frac{2|\lambda_n|^{3/2}}{3}\right)=0.
\label{zeros}
\end{equation}
These zeros can easily be found numerically. The first few are given
approximately by $\lambda_0=0,~\lambda_1=-2.338,~\lambda_2=-4.088,~\lambda_3=
-5.521,~\lambda_4=-6.787,~\lambda_5=-7.945,~\lambda_6=-9.023.$

In order to find the  the constants $C_n$ in the solution (\ref{solution}) we
notice that the basic equation (\ref{1}) is of third order, and hence does
not lead to orthogonal functions. Instead we notice that from (\ref{B-A})
the Airy function satisfies the second order differential equation,
\begin{equation}
\partial^2_x A(x)=xA(x),
\label{differential}
\end{equation}
for which standard methods are applicable: We define the auxiliary function
\begin{equation}
M(x,t)=\sum_{n=0}^\infty C_n e^{\lambda_n t}A(x+\lambda_n),
\label{expansion}
\end{equation}
and we can then easily show from the second order differential
equation (\ref{differential}) that the functions $A(x+\lambda_n)$ are
orthogonal,
and hence $C_n$ can be obtained in terms of an integral over the initial
function $M(x,0)$ times $A(x+\lambda_n)$. Since, however, we also have
\begin{equation}
\partial_x^2~M(x,t)=-N(x,t)
\label{v-u}
\end{equation}
from (\ref{solution}), the initial function $M(x,0)$ can be expressed in terms
of the corresponding initial function $N(x,0)$, giving the result
\begin{eqnarray}
C_n&=&\frac{1}{I_n}~\int_0^\infty dx~A(x+\lambda_n)\nonumber \\
&\times &\left[\int_x^\infty~dx'~
(x-x')N(x',0)-C_0~A(x)\right],
\label{cm}
\end{eqnarray}
where
\begin{eqnarray}
&&I_n=\int_0^\infty dx A(x+\lambda_n)^2=A'(\lambda_n)^2,\nonumber \\
&&{\rm and}~~C_0=\frac{-1}
{A(0)}\int_0^\infty dxxN(x,0).
\end{eqnarray}
This then allows us to write down the solution $N(x,t)$ for any given initial
function $N(x,0)$.
A characteristic feature of this solution is that the mean value $\langle x \rangle$
saturates for large times. Using the solution one finds
\begin{equation}
\langle x \rangle =\frac{\int_0^\infty x N(x,t)}{\int_0^\infty
  N(x,t)}\mathop{\longrightarrow}_{t\rightarrow \infty}\frac{3^{2/3}\Gamma (4/3)}{\Gamma (2/3)},
\label{eq15}
\end{equation}
which means that $\langle\tilde x\rangle\rightarrow 2.096(D/f)^{1/3}$.
We also find that the dispersion saturates, 
$ \langle x^2- \langle x \rangle^2 \rangle \rightarrow 0.6074
\langle x \rangle^2$, in marked contrast to pure diffusion, 
where the dispersion increases without bound.
The time development of $N(x,t)$ is governed by the Boltzmann-like factors
$\exp (\lambda_nt)$. In Fig. 1 we have shown by an example that these factors
can give rise to
additional secondary bumps, which ultimately disappear for larger times. It
would be interesting to find actual examples where these structures are
observed experimentally.
\begin{figure}[htbp]
  \begin{center}
    \epsfig{width=.4\textwidth,file=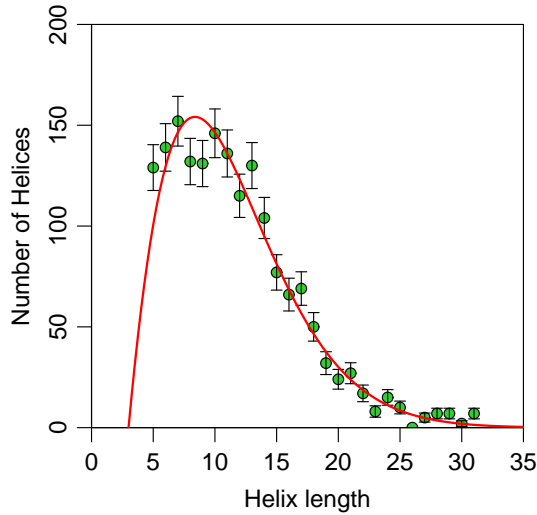}
    \caption{The black dots represent the length distribution
of $\alpha$-helices from 299 high resolution protein structures
with low homology \cite{alpha}. The error bars are estimated as
the square root of the numbers. The full curve is the stationary
distribution (i.e. infinite time limit) obtained 
from eq. (\ref{first}) plotted versus the helix
length measured in terms the number of amino acids. The ratio
between the diffusion constant $D$ and the fragmentation rate $f$
resulting in the best fit to the data is $\left(\frac{D}{f}\right)^{1/3}\approx 6.1$
number of amino acids. Note that $N(x)$ vanishes for
a helix length equal to 3, as a $\alpha$-helix needs one turn to be 
identified, see \cite{foot}.} 
    \label{fig:1}
  \end{center}
\end{figure}

We now describe two widely different application
of the suggested interplay between gradual diffusion
and sudden fragmentation.
The examples are selected from the constraint imposed not so
much by the generic nature of the two processes, but also
by the requirement that random merging of structures
should be insignificant. Further, the two examples are taken
from cases where it is reasonable to assume that fragmentation
occurs uniformly on the size axis $x$, as implicitly assumed in
the basic eq. (\ref{first}) of the process.
The two experimental data sets behind our studies are respectively
the length distribution of $\alpha$-helices in proteins, and the size
distribution of ice crystals in ice sheets. 
In the former case we will be able to extract the ratio $\frac D f$ by
performing a least squares fit of the stationary state solution and in the
latter case we do an exponential fit of $\langle x\rangle(t)$ using the two leading
terms in the solution, $\frac{\langle
  x\rangle_\infty}{1+\left(\frac{\langle x\rangle_\infty}{\langle x
      \rangle_0}-1\right)e^{-(t-t_0)/\tau}}$. We then extract both $D$
and $f$ from (\ref{eq2}) and (\ref{eq15}) and by noting that the characteristic time
$\tau$ is related to the fragmentation by the second largest eigenvalue
$\lambda_1$, we obtain $\tau=-1/(\lambda_1 f)$ \cite{is}. 
\begin{figure}[htbp]
  \begin{center}
    \epsfig{width=.43\textwidth,file=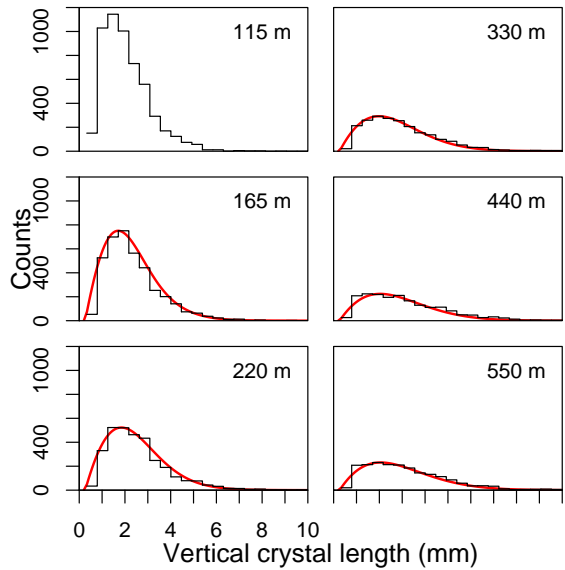}
  \end{center}
  \caption{Distributions of ice crystal sizes at
    depths 115m, 165m, 220m, 330m, 440m and 550m. The crystal size is
    defined as the vertical extension of the individual
    crystals.  The ragged curves are the measured histograms and the smooth
lines are the temporal evolution predicted by eq. (\ref{first}) starting from
    the initial distribution at 115m. The total counts of ice crystals
decreases with depth (due to the overall increase of sizes) until the steady state
is reached. The values of the diffusion constant and the fragmentation rate
used in these plots are $D = 2.8 \cdot 10^{-3} {\rm mm}^2/{\rm year}$ and
$f = 3.6 \cdot 10^{-4} /{\rm cm \cdot year}$. Note that the
    distributions vanish at a small, finite value to account for the
    experimental bias; see \cite{is}.}
  \label{fig:2}
\end{figure}

The length distribution of $\alpha$-helices \cite{alpha} is taken from
a database of 299 high resolution structures with low homology
extracted from the PDB (Protein Data Bank) \cite{alpha}.
Secondary structure have been
calculated on the basis of backbone hydrogen bonding \cite{DSSP}
and in total 2101 helices are used
\cite{alpha}.
Helices of lengths between 5 and 31 amino acids have been counted resulting
in the distribution shown as circles in Fig. \ref{fig:1}.
Clearly, the distribution exhibits a maximum around a length of 7 amino acids
followed by a long tail. 
The data do not fit well any simple 
statistical distribution and a polynomial fit requires at least a fourth order expression to
become acceptable \cite{alpha}. We consider this distribution as the
result of an 'infinitely' long evolutionary process where
diffusive growth competes with intermittent fragmentation. The
stationary solution to eq. (\ref{first}) is defined by one free
parameter (the ratio $D/f$) and it nicely reproduces the overall
rise and decline of the observed helix frequency with length \cite{foot}.
There are several points to make in this connection:
1) Considering ensembles of $\alpha$-helices, one
effectively randomizes the particular evolutionary advantages
of growing or shrinking any particular $\alpha$-helix in any particular
protein. Thus whereas a particular change may have a purpose, 
then the overall process is probably well represented by 
random domain adjustments and occasional fragmentation.
2) Though $D$ and $f$ have essentially the same molecular origin,
simple energy arguments suggest that a point mutation in the middle
of the helix to a residue not forming hydrogen bonds has a statistical
weight of $\approx 1/400$ relative to the weight of growing/shrinking
the helix \cite{doig}. This corresponds to the ratio $(D/f)^{1/3}\approx 7.4$ in
reasonable agreement with our result. 
3) That in principle other processes also acts to limit the maximum length
of $\alpha$-helices, as f.ex. the total size of proteins
limits the maximal length of their sub elements. This maximum length
will only influence the distribution of the very long $\alpha$-helices.
Overall we see the resulting distribution of $\alpha$-helices
as an ensemble average, revealing a process that leaves a statistical
stationary distribution of structural elements in much the same way
as binding energies between amino acids tend to influence the frequencies
of neighboring relations between them in an ensemble of proteins \cite{Miyazawa}.

For the ice crystal distribution, we apply data from the
North Greenland Ice Core Project (NorthGRIP) which provides
paleoclimatic information back to at least 115 kyr before present
(B.P.) \cite{DDJ}. Each year, precipitation on the ice sheet covers it
with a new layer of snow, which gradually transforms into ice crystals as the layer sinks
into the ice sheet. The size distribution of ice crystals has been
measured at selected depths in the upper 880 m of the NorthGRIP ice
core \cite{AS1}, which cover a time span of 5300 years \cite{SJJ}.
Fig. \ref{fig:2} shows size distributions of ice crystals at selected depths
down to 550 m. The crystal size $x$ is defined as the
vertical extent of a crystal \cite{comment}. 
Applying instead the horizontal extent of the crystal yields the same distribution
curve thus supporting our assumption of using the linear size in 
our formalism.

Each distribution exhibits a
pronounced peak, indicating a typical crystal size at each depth,
followed by a decaying tail of relatively large
crystals. The mean size becomes
larger with depth and thus time until it saturates \cite{TT96,LJ98}.
The
distributions gradually change with time toward a universal curve,
indicating a common underlying physical process in the formation of
crystals. We have identified this process as an interplay between the
fragmentation of the crystals and the diffusion of their grain
boundaries and is thus described by our general framework.

In this Letter we have presented a general evolutionary scenario 
for the dynamics of objects subjected to a competition between
diffusion and fragmentation. The process results in an evolution
equation of the density $N(x,t)$ of objects of linear size $x$. In the
stationary limit, this equation predicts a ``universal" distribution
curve $N(x)$ determined only by the ratio $\frac{D}{f}$. This
curve can be described exactly in terms of Bessel function leading to
a power law uprise for small $x$ followed by a tail of
a stretched exponential form for large $x$ ($exp(-x^{\frac{3}{2}})$). 
We speculate that this general
competing principle might be the relevant mechanism in many other
systems of nature and may result in distribution functions of
a similar type.

We are grateful to Mogens Levinsen and Anders Svensson for
discussions.


\begin{thebibliography}{10}
\bibitem{is} J. Mathiesen, J.
 Ferkinghoff-Borg, M.H. Jensen, M. Levinsen, P. Olesen,  D. Dahl-Jensen and A. Svensson, arXiv:physics/0310087.
\bibitem{flyvbjerg} H. Flyvbjerg, T. E Holy and S. Leibler, Phys. Rev. Lett. {\bf
    73}, 2372 (1994).
\bibitem{flyvnote}
The model proposed in \cite{flyvbjerg} for the
stochastic dynamics of microtubules focuses on transitions from the
growing to the shrinking state of microtubules and contains a fragmentation term
similar to the one in eq. (\ref{first}). However, their work considers
an open system and the model is therefore not mass conserving.
\bibitem{watson} 
G. N. Watson, {\it A treatise on the Theory of Bessel
Functions}, Cambridge University Press (1944).
\bibitem{alpha}
Penel, S., Morisson, R. G., Mortishire-Smith, R. J. and Doig, A. J.
J. Mol. Biol., 293, 1211-1219 (1999).
\bibitem{DSSP}
Kabsch, W. and Sander, C.,  Biopolymers {\bf 22}, 2577-2637 (1983).
\bibitem{foot}
The boundary condition (8) has
been replaced by $N(x_0,t)=0$, where $x_0 = 3$ in Fig. 2. In
eq. (1) $x$ is replaced by $x-x_0$, since at least four amino
acids are required to form an $\alpha$-helix, i.e. $x_0+1 = 4$.
\bibitem{doig} 
A. J. Doig, A. Chakrabartty, T. M. Klingler and R. Baldwin,
Biochemistry {\bf 33}, 3369 (1994).

\bibitem{Miyazawa}
S. Miyazawa and R.L. Jernigen, Macromolecules {\bf 18}, 534 (1985); 
S. Miyazawa and R.L. Jernigen, J. Mol. Biol. {\bf 256}, 623 (1996)
\bibitem{DDJ} D. Dahl-Jensen, N. Gundestrup, H. Miller,
  O. Watanabe, S.J. Johnsen, J.P Steffensen, H.B. Clausen,
  A. Svensson, and L.B. Larsen,
  {\it Annals of Glaciology} {\bf 35}, 1-5, 2002.
\bibitem{AS1} A. Svensson, K.G. Schmidt, D. Dahl-Jensen,
  S.J. Johnsen, Y. Wang, J. Kipfstuhl, and T. Thorsteinsson,
  Annals of Glaciology {\bf 37} (2003).
\bibitem{SJJ}  S.J. Johnsen, D. Dahl-Jensen, N. Gundestrup,
  J.P. Steffensen, H.B. Clausen, H. Miller, V. Masson-Delmotte,
  A.E. Sveinbj\"orndottir, and J. White, Journal of
  Quaternary Science {\bf 16} (4), 299-307, 2001.
\bibitem{comment}
The model, eq (\ref{first}), can be generalized by changing the
fragmentation rate to be proportional to $x^\beta$ to account for
energy/surface arguments. The stationary distribution can also be
solved in these cases. For $\beta=2$ the solution will be 
$A(x)\sim \sqrt{x} K_{1/4}(x^2/2)$ instead of eq. (\ref{bessel}).
However, this expression provides a poor fit to the experimental 
data for the ice crystals.
\bibitem{TT96} T. Thorsteinsson,
  Berichte zur Polarforschung {\bf 205}, 1-146 (1996).
\bibitem{LJ98} L. Jun, T.H. Jacka, and V. Morgan,
Ann. Glaciology {\bf 27}, 343-348 (1998).
\end{thebibliography}
\end{document}